\documentclass[sigconf, nonacm]{acmart}
\AtBeginDocument{%
  }

\setcopyright{acmlicensed}
\copyrightyear{2026}
\acmYear{2026}
\acmDOI{XXXXXXX.XXXXXXX}
\acmConference[WWW]{The ACM Web Conference}{April 13--17,
  2026}{Dubai, UAE}
\acmISBN{978-1-4503-XXXX-X/2018/06}

\newcommand{\best}[1]{\textbf{#1}}
\newcommand{\second}[1]{\uline{#1}}

\newcommand{\methodblock}[2]{
    \textbf{\colorbox{methodColor}{#1}}\\[0.3em]
    {\small
    \begin{minipage}[t]{\linewidth}
    \begin{enumerate}[leftmargin=*, itemsep=2pt]
        #2
    \end{enumerate}
    \end{minipage}
    }
}

\usepackage{ulem}
\usepackage{xcolor}       % For text coloring
\usepackage{graphicx}     % For resizebox
\usepackage{booktabs}     % For toprule, midrule, bottomrule
\usepackage{enumitem}     % For custom list environments
\usepackage{tcolorbox}
\tcbuselibrary{skins, breakable}
\usepackage{microtype}
\usepackage{subcaption}
\usepackage{multirow} % for main experiment
\usepackage{siunitx}
\usepackage{arydshln}
\usepackage[utf8]{inputenc} % if you compile with pdfLaTeX
\DeclareUnicodeCharacter{202F}{\,} % map U+202F to a thin space
\begin{document}


% \title{Doc2Query$++$: Topic-Coverage Document Expansion via Few-shot Query Generation and Dual-Index Fusion for Information Retrieval}
\title{Doc2Query$++$: Topic-Coverage based Document Expansion and its Application to Dense Retrieval via Dual-Index Fusion}

\author{Tzu-Lin Kuo}
\authornote{Both authors contributed equally to this research.}
\email{r12922050@ntu.edu.tw}
\orcid{1234-5678-9012}
\affiliation{%
  \institution{National Taiwan University}
  \city{Taipei}
  \country{Taiwan}
}
\author{Wei-Ning Chiu}
\authornotemark[1]
\email{r12922219@ntu.edu.tw}
\affiliation{%
  \institution{National Taiwan University}
  \city{Taipei}
  \country{Taiwan}
}

\author{Wei-Yun Ma}
\email{ma@iis.sinica.edu.tw}
\affiliation{%
  \institution{Academia Sinica}
  \city{Taipei}
  \country{Taiwan}
}
\author{Pu-Jen Cheng}
\email{pjcheng@csie.ntu.edu.tw}
\affiliation{%
  \institution{National Taiwan University}
  \city{Taipei}
  \country{Taiwan}
}
\renewcommand{\shortauthors}{Kuo et al.}

\begin{abstract}
Document expansion (DE) via query generation tackles vocabulary mismatch in sparse retrieval, yet faces limitations: uncontrolled generation producing hallucinated or redundant queries with low diversity; poor generalization from in-domain training (e.g., MS MARCO) to out-of-domain data like BEIR; and noise from concatenation harming dense retrieval. While Large Language Models (LLMs) enable cross-domain query generation, basic prompting lacks control, and taxonomy-based methods rely on domain-specific structures, limiting applicability. To address these challenges, we introduce Doc2Query$++$, a DE framework that structures query generation by first inferring a document's latent topics via unsupervised topic modeling for cross-domain applicability, then using hybrid keyword selection to create a diverse and relevant keyword set per document. This guides LLM not only to leverage keywords, which ensure comprehensive topic representation, but also to reduce redundancy through diverse, relevant terms. To prevent noise from query appending in dense retrieval, we propose Dual-Index Fusion strategy that isolates text and query signals, boosting performance in dense settings. Extensive experiments show Doc2Query++ significantly outperforms state-of-the-art baselines, achieving substantial gains in MAP, nDCG@10 and Recall@100 across diverse datasets on both sparse and dense retrieval.
\end{abstract}


% \begin{CCSXML}






\maketitle

\section{Introduction}
% !TeX root = ../main.tex
One of the central challenges in information retrieval (IR) is the vocabulary mismatch problem—users often express information needs using terms that differ from those in relevant documents, reducing recall and hindering performance in open-domain settings \cite{furnas1987vocabulary,zhao2010term}. Traditional strategies to mitigate this include query expansion, which enriches user queries with related terms from feedback or external resources, and document expansion (DE), which augments documents themselves with predicted queries or contextual terms prior to indexing. While query expansion dominates due to its flexibility—no re-indexing required \cite{nogueira2019document} - DE offers a key advantage: it leverages full-document context for more semantically grounded representations, potentially yielding stable gains over fragile, query-dependent expansions \cite{billerbeck2004questioning, 10.1145/1645953.1646059, 10.1007/978-3-642-00958-7_57}.

Recent neural advances have revitalized DE through query generation (QG). Doc2Query \cite{nogueira2019document} trains a sequence-to-sequence model on in-domain data (e.g., MS MARCO) to generate synthetic queries per document, appending them for indexing and outperforming baselines like RM3 in sparse retrieval. Doc2Query$--$ \cite{gospodinov2023doc2query} refines this by filtering low-relevance queries, reducing noise and index size. However, these methods face several limitations: (1) uncontrolled generation often produces hallucinated, redundant queries with low diversity and incomplete topic coverage; (2) they generalize poorly from in-domain training to out-of-domain (OOD) benchmarks like BEIR \cite{thakur2021beir, mansour2024revisiting}; (3) direct query appending injects semantic noise that degrades dense retrieval performance, as stronger encoders become sensitive to such distortions \cite{weller2023generative}. 

To address the generalizability issues inherent in prior neural DE methods, we explore prompt-based QG via Large Language Models (LLMs) \cite{dai2022promptagator, jeronymo2023inpars, sannigrahi2024synthetic, chaudhary2023s}, which offer promising cross-domain capabilities \cite{brown2020language, sanh2021multitask}. Yet, prompt-based QG still suffers from uncontrolled output, leading to similar issues of redundancy and omitted document facets \cite{kang2025improving}. A notable exception is CCQGen \cite{kang2025improving}, which incorporates concept coverage by inferring topics from a predefined academic taxonomy, extracting keywords, and iteratively generating queries to address uncovered concepts—demonstrating gains in scientific retrieval. However, CCQGen's reliance on domain-specific taxonomies limits its applicability to open-domain IR, where such structures are unavailable or impractical.

\definecolor{loanColor}{RGB}{30, 144, 255}     % DodgerBlue
\definecolor{investColor}{RGB}{34, 139, 34}     % ForestGreen
\definecolor{disposalColor}{RGB}{220,20,60} % crimson for
\definecolor{methodColor}{RGB}{220, 230, 250}

\begin{table}[!htbp]
    \centering
    \renewcommand{\arraystretch}{0.9}
    \resizebox{1\linewidth}{!}{
    \begin{tabular}{p{0.47\linewidth} | p{0.47\linewidth}} \toprule

    % --- DOCUMENT SECTION (Spanning two columns) ---
    \multicolumn{2}{c}{
        \parbox{0.95\linewidth}{
            \textbf{Document} (\textbf{Topic1:} \textcolor{loanColor}{UK Student Loan},
            \textbf{Topic2:} \textcolor{investColor}{Debt \& Investment Strategy},
            \textbf{Topic3:} \textcolor{disposalColor}{Asset Disposal})\\ \rule{\linewidth}{0.4pt}
            \small
            I'm surprised no one has picked up on this, but the \textcolor{loanColor}{student loan is an exception to the rule}. It's \textcolor{loanColor}{inflation bound}, you only have to \textcolor{loanColor}{pay it back as a percentage of your salary if you earn over £15k}… you \textcolor{loanColor}{don't have to pay it if you lose your job}, and it \textcolor{loanColor}{doesn't affect your ability to get credit}.
            My advice… is: if you have any \textcolor{disposalColor}{shares that have lost more than 10 \%} since you bought them… \textcolor{disposalColor}{sell them and pay off your debts with those}. The rest is down to you – are they making more than 10\% a year? If they are, don't sell them. If your \textcolor{investColor}{dividends are covering your payments}, carry on as you are.
        }
    } \\ \midrule

    % --- GENERATED QUERIES TITLE (Spanning two columns) ---
    \multicolumn{2}{c}{\textbf{Generated Queries}} \\ \midrule

    % --- FIRST LEVEL: Doc2Query and Doc2Query-- (Side-by-side) ---
    \begin{minipage}[t]{\linewidth}
        \methodblock{Doc2Query}{
            \item[$q^1$] can you pay out \textcolor{loanColor}{student loan} with a job?
            \item[$q^2$] can you borrow from \textcolor{loanColor}{student loan}?
            \item[$q^3$] do you have to pay \textcolor{loanColor}{student loan} back?
        }
    \end{minipage} &
    \begin{minipage}[t]{\linewidth}
        \methodblock{Doc2Query$--$}{
            \item[$q^1$] can i pay my \textcolor{loanColor}{student loan} if i lose my job?
            \item[$q^2$] do \textcolor{loanColor}{student loan} have to be paid back if you lose your job?
            \item[$q^3$] can you pay a \textcolor{loanColor}{student loan} back?
        }
    \end{minipage} \\ \midrule

    % --- SECOND LEVEL: Doc2Query++ (Spanning two columns) ---
    \multicolumn{2}{c}{
        \parbox{0.95\linewidth}{
            \methodblock{Doc2Query++ (Ours)}{
                \item[$q^1$] How do \textcolor{loanColor}{income-based repayment plans} affect \textcolor{loanColor}{student loan} burdens for individuals earning over £15,000 a year?
                \item[$q^2$] How do \textcolor{investColor}{dividend payments impact debt repayment strategies} for investors with fluctuating income streams?
                \item[$q^3$] What are the key factors to consider when deciding whether to \textcolor{disposalColor}{sell underperforming shares to pay off debts}?
            }
        }
    } \\ \bottomrule
    \end{tabular}
    }
    \caption{Comparison of generated queries across different methods for the same input document, with per‐method topic-coverage.}
    \label{tab:query-examples}
\end{table}

To overcome these challenges, we introduce Doc2Query$++$, a unified DE framework that ensures comprehensive topic coverage and balanced term augmentation for effective sparse and dense retrieval. Drawing inspiration from CCQGen's control mechanisms, we replace fixed taxonomies with unsupervised topic modeling (via BERTopic \cite{grootendorst2022bertopic}) to infer latent document topics in a domain-agnostic manner. We then construct a hybrid keyword set—merging topic-derived terms for semantic novelty with document-specific phrases for grounding—and prompt an LLM to generate multiple queries in a single pass that collectively span all topics while incorporating diverse, relevant keywords to minimize lexical overlap. As shown in Table \ref{tab:query-examples}, this yields queries that cover underrepresented facets (e.g., "Debt \& Investment Strategy"), unlike prior methods that focus narrowly on dominant themes. Furthermore, we extend DE to dense retrieval—where naive appending harms precision—via a Dual-Index Fusion strategy: we maintain separate indices for original document embeddings and generated query embeddings, aggregating signals at inference time with tunable weighting. This isolates noise while harnessing expansion benefits, transforming DE from detrimental to advantageous in dense settings. Our key contributions are as follows:
\begin{itemize}
    \item \textbf{Topic-aware document expansion}: We incorporate explicit topic coverage into neural DE, ensuring generated queries comprehensively represent each document's semantic structure.
    \item \textbf{Unsupervised topic modeling for cross-domain applicability}: Leveraging BERTopic, we uncover latent structures without pre-defined taxonomies, enabling adaptation across diverse domains.
    \item \textbf{Hybrid keyword selection with LLM prompting}: We combine topic-level and document-level signals to guide query generation, balancing lexical specificity and novelty.
    \item \textbf{Dense retrieval integration}: Addressing degraded dense retrieval performance caused by noise injected by direct query appending \cite{weller2023generative}, our Dual-Index Fusion isolates and fuses query/text embeddings, boosting both precision and recall.
\end{itemize}

Extensive experiments on BEIR subsets of various domains show Doc2Query$++$ outperforms baselines in MAP, nDCG@10 and Recall@100 in both sparse and dense settings, validating its efficacy across diverse domains.

\section{Related Works}
% !TeX root = ../main.tex
\subsection{Document Expansion}
Document expansion (DE) addresses vocabulary mismatch by enhancing documents with relevant terms before indexing. For a document collection $D = \{d_1, d_2, \dots, d_n\}$, an expansion method generates synthetic queries $Q_i = \{q_{i1}, q_{i2}, \dots, q_{im}\}$ for each document $d_i$. These queries are appended to the original document, forming an expanded document $d_i' = d_i \cup Q_i$. The resulting expanded corpus $D' = \{d_1', d_2', \dots, d_n'\}$ is then indexed and used for search.

\subsubsection{Early Approaches to Document Expansion}
Early document expansion methods relied on external signals and statistical models. Heuristic approaches leveraged terms from web structures like inbound anchor text and URLs \cite{craswell2001effective, eiron2003analysis, westerveld2002retrieving}, or from user behavior data like historical search queries \cite{pickens2010reverted, scholer2004query}. A parallel approach used statistical language models to algorithmically generate relevant expansion terms \cite{efron2012improving, tao2006language}.

\subsubsection{Neural Model Approaches to Document Expansion}
The advent of neural models has significantly advanced document expansion techniques, shifting from heuristic-based to learning-based methods. Doc2Query \cite{nogueira2019document} used a sequence-to-sequence transformer to produce queries, resulting in modest effectiveness gains over the BM25 baseline. It was later shown that replacing this simple transformer with the much larger T5 model—an approach known as DocT5Query—yielded substantially greater improvements \cite{nogueira2019doc2query}. While the latest large language models like ChatGPT have also been explored for this task, it remains unclear if they outperform established T5-based methods \cite{weller2023generative}. A key challenge with these generative models is their tendency to include irrelevant or hallucinated terms. To address this, Doc2Query$--$ \cite{gospodinov2023doc2query} adds a filtering mechanism that uses a relevance model to score generated query-document pairs and retain only the queries above a set threshold. However, these methods still lack focus on the nature of generated queries, presenting three key challenges: (1) uncontrolled generation results in redundant concepts, uncovered sections, and high lexical overlap, restricting coverage; (2) document expansion in dense retrieval remains underexplored, with direct concatenation impairing performance \cite{weller2023generative}; (3) as learning-based approaches trained on in-domain data, they struggle to generalize to out-domain datasets \cite{mansour2024revisiting}. Our Doc2Query$++$ addresses these by guiding query generation through topic modeling, keyword extraction, and Dual-Index Fusion.

\subsection{Prompt-based Query Generation}
The rise of Large Language Models (LLMs), with their robust language understanding, has enabled prompt-based generation of human-like, contextually relevant text without costly task-specific training. Early methods like InPars \cite{bonifacio2022inpars} prompt models with documents to create relevant queries for unsupervised IR datasets, while Promptagator \cite{dai2022promptagator} uses few-shot learning with document-query examples to shape output styles. Prompting has evolved, with techniques like pairwise approach, which generates relevant queries followed by less-relevant "hard negatives" to boost retriever training \cite{chaudhary2023s}. Despite these advancements, existing prompting struggles to ensure comprehensive coverage of a document’s distinct concepts. CCQGen \cite{kang2025improving} addresses this in scientific contexts by integrating a control mechanism with a pre-defined academic topic taxonomy, guiding query generation for broader and more diverse concept coverage. However, CCQGen’s fixed taxonomy restricts cross-domain applicability, requiring curation for new fields. Building on our Doc2Query$++$ framework, this limitation is addressed through unsupervised topic modeling, enabling adaptive, domain-agnostic query generation across diverse contexts.

\section{Doc2Query$++$}
% !TeX root = ../main.tex
% \begin{figure}[!htbp]
%     \centering
%     \includegraphics[width=\paperwidth]{figures/method_overview.pdf}
%     \caption{Method Overview}
%     \label{fig:method-overview}
% \end{figure}

\begin{figure*}[!htbp]
  \centering
  \includegraphics[width=\textwidth]{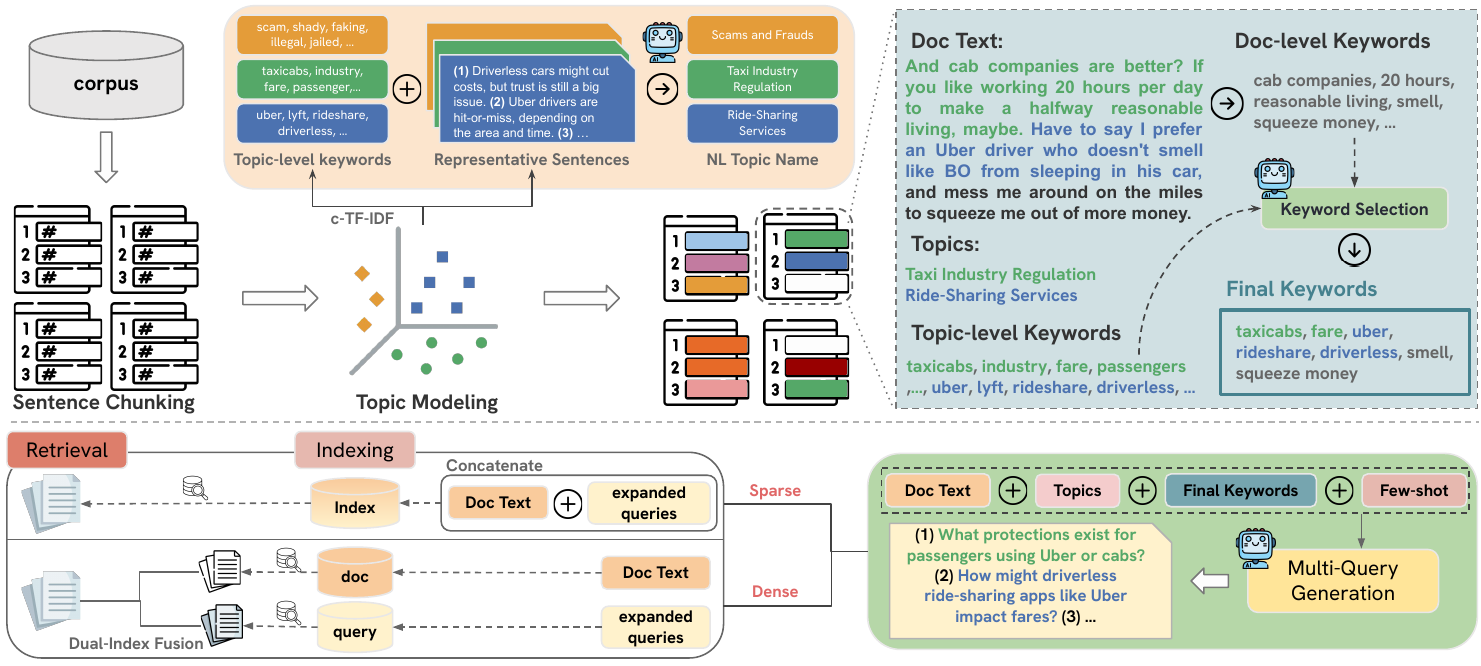}
  \caption{Overview of the Doc2Query$++$ framework. The model infers latent topics via BERTopic, extracts topic- and document-level keywords, and uses them to guide LLM-based query generation. Generated queries are appended for sparse retrieval and fused through Dual-Index Fusion for dense retrieval.}
  \label{fig:method_overview}
\end{figure*}
We introduce \textbf{Doc2Query$++$}, a unified DE framework designed to generate queries that satisfy two critical objectives: (1) \textbf{Comprehensive Topic Coverage}: Generated queries should collectively span all thematic areas of a document, ensuring complete topical representation. (2) \textbf{Balanced Term Augmentation}: Queries must reinforce each document’s existing key terms and introduce new, semantically relevant vocabulary to bridge lexical gaps between documents and potential user queries. Doc2Query$++$ achieves these goals through a structured, four-stage pipeline. Below, we provide a high-level overview of the four stages before detailing their implementation:

\begin{enumerate}
    \item \textbf{Sentence-level Topic Modeling \& Representation Refinement}: We first perform unsupervised topic modeling by clustering sentence embeddings for each document using BERTopic, identifying its latent topic set and computing initial topic representations and centroids. We then use an LLM to refine these topic representations into concise, human-readable labels via few-shot prompting.
    \item \textbf{Keyword Extraction and Selection}: We obtain two complementary keyword sets—one capturing high-level themes inferred by topic modeling and one extracting salient in-text phrases directly from each document. After merging these keyword pools into a unified candidate list, we prompt an LLM to select a focused subset that both represents the document’s core topics and injects novel semantic cues.
    \item \textbf{Topic-Coverage based Query Generation}: Leveraging the topic names and focused keyword set from previous stages, we instruct an LLM to generate multiple queries per document in one pass. After generating the query sets, we apply them differently for sparse and dense retrieval methods.
    \item \textbf{Sparse and Dense Retrieval with Expansion and Dual-Index Fusion}: For sparse retrieval, we follow standard expansion by appending generated queries to document text. For dense retrieval, since naive appending injects semantic noise and harms dense retrievers \cite{weller2023generative}, we address this by proposing the Dual-Index Fusion strategy: maintaining separate indices for original document embeddings and query embeddings, aggregating signals via max-pooling, and fusing scores with a tunable weight.
\end{enumerate}
An overview of this pipeline is illustrated in Figure~\ref{fig:method_overview}, providing a visual summary of the topic modeling, keyword selection, query generation, and Dual-index Fusion components described above.

\subsection{Topic Modeling and Representation Refinement}

\subsubsection{Sentence-level Topic Modeling through BERTopic}
To capture each document’s latent semantic structure without manual tuning, we adopt an embedding-based clustering approach instantiated by BERTopic. Unlike traditional lexical-based topic-modeling methods \cite{blei2003lda, hofmann2013probabilistic, papadimitriou1998latent}, BERTopic operates directly on contextualized text embeddings and dynamically infers the optimal number of topics for a given corpus. Since sentences are the natural minimal units that represent coherent topical content, we segment each document at the sentence level \footnote{We use the \texttt{sent\_tokenize} function from NLTK~\cite{bird-loper-2004-nltk}.} before modeling. 

Formally, we consider a corpus $\mathcal{D}=\{d_1, \dots, d_N\}$, where each document $d$ is a set of sentences $S_d$. We first encode every sentence $s$ from the overall sentence set $\mathcal{S} = \bigcup_{d \in \mathcal{D}} S_d$ into an $m$-dimensional embedding $\mathbf{z}_s = f_{\mathrm{enc}}(s)$ using a pre-trained, domain-specific SBERT encoder $f_{\mathrm{enc}}: \mathcal{S} \to \mathbb{R}^m$ \cite{reimers-2019-sentence-bert}. Next, BERTopic clusters the sentence embeddings $\{\mathbf{z}_s\}_{s \in \mathcal{S}}$ via HDBSCAN \cite{mcinnes2017accelerated}, a density-based algorithm that automatically determines the number of topic clusters, say $C$. Each cluster $j \in \{1, \dots, C\}$ is represented by a centroid vector $\boldsymbol{\mu}_j$. We then assign each sentence embedding $\mathbf{z}_s$ to its nearest topic cluster $\phi(\mathbf{z}_s)$ using L2 distance:
\begin{equation*}
\label{eq:topic-assignment}
\phi(\mathbf{z}_s) = \arg\min_{j=1,\dots,C} \|\mathbf{z}_s - \boldsymbol{\mu}_j\|_2,
\end{equation*}
excluding any outlier embeddings identified by HDBSCAN. Finally, the topic representation for a document $d$ is the union of topics assigned to its sentences:
\begin{equation*}
T(d) = \bigcup_{s \in S_d} \phi(\mathbf{z}_s).
\end{equation*}
This unsupervised approach allows us to model a document's latent topics by leveraging contextual embeddings, ensuring generalizability across diverse domains.

\subsubsection{Fine-Tuning Topic Representations via LLM}
\label{sec:topic-refine-llm}
Since unsupervised clustering yields topic clusters without explicit labels, we refine them into coherent, human-readable names to better guide the LLM in generating queries that span all specified topics. We achieve this using BERTopic's class-based TF-IDF (c-TF-IDF) implementation~\cite{grootendorst2022bertopic}, where sentences assigned to each cluster are concatenated into a pseudo-document. Term frequency is computed within each pseudo-document, with inverse document frequency measured across the cluster collection, producing weighted scores that highlight terms both representative of and distinctive to each topic. From these, we select the top-$M$ keywords as provisional descriptors. Next, we identify the top-$L$ representative sentences per topic—those whose embeddings are closest to the topic centroid—and feed them, along with the corresponding c-TF-IDF keywords, into the LLM via few-shot prompting (see Appendix \ref{topic-name-refinement-prompt}) to generate concise, natural-language topic names. In our experiments, we adopt BERTopic's default settings ($M=10$, $L=3$), which yield robust and coherent results.

\subsection{Keyword Extraction and Selection}
\label{sec:keywords_extract_select}
Our initial method directly employed the top-10 c-TF–IDF keywords from each document’s inferred topic set to guide query generation. However, this approach introduced substantial noise: many cluster-level keywords—while salient at the corpus level—were irrelevant to individual documents. For example, a document on simple moving average crossover strategies for equity trading shares generic financial terms like “market”, “trend”, and “volatility” with cryptocurrency trading documents, so unsupervised clustering can group it under a crypto-focused topic. This yields extraneous keywords such as “blockchain", “bitcoin”, “mining”, and “smart contract” for the document, but none of which relate to moving averages or stock markets, which steers the LLM away from the document’s core content.

To address this, we complement topic-level keywords \(K^{\mathrm{topic}}_d\) with doc\-ument-level extraction using KeyBERT \cite{grootendorst2020keybert}, whose embedding-based mechanism aligns with BERTopic by leveraging SBERT representations to identify semantically meaningful phrases beyond raw token counts. Specifically, we apply KeyBERT to  encode each document and its candidate n-grams, ranking them by cosine similarity to the document embedding and generating a top-20 n-grams (n=1–3) per document; to promote diversity and minimize redundancy, we then use Maximal Marginal Relevance (MMR, \(\lambda=0.7\)), yielding our document-level keywords \(K^{\mathrm{doc}}_d\). Next, let \(K^{\mathrm{topic}}_d\) and \(K^{\mathrm{doc}}_d\) denote the topic- and document-level keywords for document \(d\), respectively. Here, \(K^{\mathrm{topic}}_d\) provides “new-term” signals for broader thematic novelty, while \(K^{\mathrm{doc}}_d\) emphasize salient in-text n-grams reflecting the document’s specific content. Inspired by \citeauthor{lin2022pretrained}'s Doc2Query analysis showing gains from both new and copied terms \cite{lin2022pretrained}, we first form a unified candidate keyword pool:
\[
K_d = K^{\mathrm{topic}}_d \;\cup\; K^{\mathrm{doc}}_d.
\]
We then instruct an LLM to select a final subset \(\widehat{K}_d \;\subset\; K_d\) (with \(\bigl|\widehat{K}_d\bigr| = 10\)) by asking it to choose the keywords that best align with the document's core theme (see Appendix \ref{keyword-selection-prompt} for the prompt). To verify that topic-only, document-only, and hybrid keyword sources each contribute uniquely to retrieval effectiveness, we conduct an ablation study in Section~\ref{llm-keyword-select-effectiveness}, demonstrating that our LLM-guided fusion outperforms either source used in isolation.

\subsection{Topic-Coverage based Query Generation}
\label{method:qg-stage}
Given a document $d$ with its inferred topic set $T_d$ and keyword list $K_d$ obtained from above stages, our goal is to generate a set of $M$ queries that collectively cover all topics in $T_d$ and incorporate relevant keywords from $K_d$ to enhance lexical diversity and relevance. Prior LLM-based QG methods \cite{dai2022promptagator, bonifacio2022inpars, jeronymo2023inpars} typically produce one query at a time without explicit mechanisms for topic coverage, while CCQGen \cite{kang2025improving} uses iterative classification to enforce coverage, which can be computationally intensive. In contrast, we leverage the LLM's autoregressive capabilities and contextual understanding to generate queries in batches, ensuring coverage in an efficient manner.

Specifically, we prompt the LLM with the document text, $T_d$, and $K_d$, using a few-shot template that includes $n$ exemplar (e.g., document-topic-keyword-multi-query tuples) demonstrating how to distribute queries across topics while weaving in keywords (see Appendix \ref{few-shot-prompt-qg} for the full prompt). To balance generation efficiency with coverage control, we generate queries in batches of size $b=3$ until reaching $M$. In qualitative assessments, we observed that single-query generation often leads to incomplete coverage by overlooking remaining topics, whereas our multi-query approach consistently spans all facets, yielding a more faithful semantic representation without the overhead of post-generation filtering or classifiers.

\subsection{Sparse and Dense Retrieval with Expansion and Dual-Index Fusion}
\label{dual-index-method}

Most prior document‐expansion methods target sparse retrieval alone \cite{nogueira2019doc2query, nogueira2019document, gospodinov2023doc2query}. Accordingly, for sparse retrieval, we also follow this standard document‐expansion approach: we concatenate each document’s generated queries to its original text, rebuild the inverted index over this expanded corpus (for example, using BM25), and then perform retrieval as usual. While this approach reliably boosts both recall and precision in sparse retrieval by introducing new terms and mitigating vocabulary mismatch, naively appending expansions degrades dense retrieval effectiveness \cite{weller2023generative} due to semantic noise injected by expanded queries, which weakens the original document embedding. To address this, we propose a Dual-Index Fusion strategy that decouples expanded queries from raw text. Specifically, let each document $d$ have $M$ expanded queries ${q_{d,1},\dots,q_{d,M}}$ generated through a QG model. We construct two separate indices: (1) text index \(\mathcal{I}_t\) of original document embeddings \(\mathbf{v}_d\) and (2) query index \(\mathcal{I}_q\) of query embeddings \(\mathbf{u}_{d,i}\), each linked back to its source document \(d\). At retrieval time, given a user query $Q$ with embedding $\mathbf{v}_Q$, we perform:

\begin{enumerate}
  \item \textbf{Text retrieval}
    We query \(\mathcal{I}_t\) to obtain the top–\(n_t\) document candidates
    \[
      \mathcal{D}_t = \{d_1^t,\dots, d_{n_t}^t\},
    \]
    where each document \(d\in\mathcal{D}_t\) is scored by
    \[
      S_t(d) \;=\; \mathrm{sim}\bigl(\mathbf{v}_Q,\mathbf{v}_d\bigr).
    \]
    This captures the similarity between the query and the original document embeddings.

  \item \textbf{Query retrieval and aggregation:}\\
  Simultaneously, we query \(\mathcal{I}_q\) to retrieve the top–\(n_q\) expanded‐query embeddings nearest to the user query vector \(\mathbf{v}_Q\).  
Let \(\mathcal{Q} = \{q_1, q_2, \dots, q_N\}\) denote all generated queries from the entire corpus, where each \(q_j\) has a dense embedding \(\mathbf{u}_j\), and is associated with a source document via a mapping function \(\mathrm{doc}(j)\).

Let
\[
\mathcal{H}_q = \{\, j_1, \dots, j_{n_q} \} \subseteq \{1, \dots, N\}
\]
be the indices of these retrieved query embeddings (global query indices).  
For each document \(d\), define the set of retrieved queries originating from \(d\) as:
\[
\mathcal{Q}_d = \{\, j \in \mathcal{H}_q \mid \mathrm{doc}(j) = d \}.
\]
For each \(j \in \mathcal{Q}_d\), compute the similarity score:
\[
S_q(d, j) = \mathrm{sim}\bigl(\mathbf{v}_Q, \mathbf{u}_j \bigr).
\]
Since multiple queries may be retrieved from the same document, we aggregate their relevance using max‐pooling:
\[
S_q(d) = 
\left\{
  \begin{aligned}
    &\max_{j \in \mathcal{Q}_d} S_q(d, j) &&\quad \text{if } \mathcal{Q}_d \neq \emptyset, \\
    &0                                   &&\quad \text{if } \mathcal{Q}_d = \emptyset.
  \end{aligned}
\right.
\]

i.e., if none of \(d\)’s expanded queries appear in the top–\(n_q\) retrieved set, then \(S_q(d) = 0\).

    % Simultaneously, we query \(\mathcal{I}_q\) to retrieve the top–\(n_q\) expanded‐query embeddings nearest to the user query vector \(\mathbf{v}_Q\).  Let 
    % \[
    %   \mathcal{H}_q = \{\,i_1,\dots,i_{n_q}\}\subseteq\{1,\dots,M\}
    % \]
    % be the indices of these retrieved embeddings.  For each document \(d\), define
    % \[
    %   \mathcal{Q}_d^+
    %     = \{\,i \mid \text{embedding }i\text{ was generated from }d\},
    %   \qquad
    %   \mathcal{Q}_d = \mathcal{H}_q \,\cap\, \mathcal{Q}_d^+.
    % \]
    % For each \(i\in\mathcal{Q}_d\), compute
    % \[
    %   S_q(d,i)
    %   = \mathrm{sim}\bigl(\mathbf{v}_Q,\mathbf{u}_{i}\bigr).
    % \]
    % Since multiple hits may map back to the same document, we collapse these into a single per‐document score via max‐pooling:
    % \[
    % S_q(d) = 
    % \left\{
    %   \begin{aligned}
    %     &\max_{i \in \mathcal{Q}_d} S_q(d, i) &&\quad \text{if } \mathcal{Q}_d \neq \emptyset, \\
    %     &0                                   &&\quad \text{if } \mathcal{Q}_d = \emptyset.
    %   \end{aligned}
    % \right.
    % \]
    
    % i.e.\ if none of \(d\)’s expanded queries appear in the top–\(n_q\) retrieval set, then \(S_q(d)=0\).

  \item \textbf{Score Fusion:}\\
    Finally, we combine text‐ and query‐based scores into a unified relevance metric:
    \[
      S(d) = (1 - \alpha)\,S_t(d) + \alpha\,S_q(d),
      \quad \alpha\in[0,1].
    \]
\end{enumerate}

Here, \(\mathrm{sim}(\cdot,\cdot)\) denotes a similarity function (e.g., dot product or cosine). Setting \(\alpha=0\) recovers pure text-based dense retrieval, while \(\alpha=1\) relies solely on query-derived signals.

This Dual-Index Fusion strategy restores and surpasses the performance lost to naive query appending by isolating semantic noise and exploiting query expansions through principled aggregation and fusion. We demonstrate its effectiveness across dense retrieval benchmarks in Section~\ref{sec:dual-index-effectiveness}.

\section{Experiments}
% !TeX root = ../main.tex
\begin{table*}[!htbp]
  \centering
  \resizebox{\linewidth}{!}{%
  \begin{tabular}{llrrr  rrr  rrr  rrr  rrr}
    \toprule
    \multicolumn{2}{c}{} 
      & \multicolumn{3}{c}{\textbf{NFCorpus}} 
      & \multicolumn{3}{c}{\textbf{SCIDOCS}} 
      & \multicolumn{3}{c}{\textbf{FiQA-2018}}
      & \multicolumn{3}{c}{\textbf{Arguana}}
      & \multicolumn{3}{c}{\textbf{Scifact}}\\
    \cmidrule(lr){3-5} \cmidrule(lr){6-8} \cmidrule(lr){9-11} \cmidrule(lr){12-14} \cmidrule(lr){15-17} 
    \textbf{Retrieval} & \textbf{Method} 
      & \textbf{MAP} & \textbf{N@10} & \textbf{R@100} 
      & \textbf{MAP} & \textbf{N@10} & \textbf{R@100} 
      & \textbf{MAP} & \textbf{N@10} & \textbf{R@100} 
      & \textbf{MAP} & \textbf{N@10} & \textbf{R@100}
      & \textbf{MAP} & \textbf{N@10} & \textbf{R@100} \\
    \midrule
    \multirow{6}{*}{Sparse} 
      & BM25           & 0.1489 & 0.3223 & 0.2438 & 0.0977 & 0.1430 & 0.3433 & 0.2038 & 0.2466 & 0.5523 & 0.2381 & 0.3437 & 0.9450 & 0.6384 & 0.6776 & 0.9198 \\
      & Doc2Query      & 0.1544 & 0.3338 & 0.2470 & \second{0.1047} & \second{0.1495} & \second{0.3592} & 0.2348 & 0.2786 & 0.5753 & 0.2408 & 0.3475 & 0.9493 & \second{0.6572} & \second{0.6927} & \best{0.9248} \\
      & Doc2Query––    & 0.1551 & 0.3334 & 0.2495 & 0.1025 & 0.1487 & 0.3586 & 0.2336 & 0.2771 & 0.5811 & 0.2321 & 0.3312 & 0.9429 & 0.6518 & 0.6893 & 0.9186 \\
      & Zero-Shot LLM  & 0.1519 & 0.3221 & 0.2451 & 0.0976 & 0.1398 & 0.3397 & 0.2381 & 0.2828 & \second{0.5881} & \second{0.2463} & \second{0.3558} & \second{0.9586} & 0.6529 & 0.6878 & 0.9217 \\
      & Few-Shot LLM   & \second{0.1580} & \second{0.3371} & \second{0.2531} & 0.0967 & 0.1406 & 0.3323 & \second{0.2436} & \second{0.2877} & 0.5864 & 0.2317 & 0.3304 & 0.9465 & 0.6513 & 0.6888 & 0.9194 \\
      & \textbf{Doc2Query++ (Ours)} & \best{0.1602} & \best{0.3415} & \best{0.2639} & \best{0.1095} & \best{0.1568} & \best{0.3749} & \best{0.2488} & \best{0.2972} & \best{0.6197} & \best{0.2487} & \best{0.3561} & \best{0.9629} & \best{0.6641} & \best{0.6945} & \best{0.9248}\\
    \midrule
    \multirow{6}{*}{Dense} 
      & Contriever     & 0.1618 & 0.3303 & 0.2951 & 0.1022 & 0.1459 & 0.3780 & 0.2748 & 0.3324 & 0.6564 & 0.2295 & 0.3304 & 0.9557 & 0.6218	& 0.6574 & \second{0.9467} \\
      & Doc2Query      & 0.1612 & 0.3401 & 0.2878 & 0.1097 & 0.1577 & 0.3697 & 0.2789 & 0.3316 & 0.6406 & 0.2373 & 0.3401 & 0.9586 & 0.6390 & 0.6785 & 0.9267\\
      & Doc2Query––    & 0.1577 & 0.3353 & 0.2823 & 0.1070 & 0.1559 & 0.3631 & 0.2818 & 0.3339 & 0.6226 & 0.2364 & 0.3387 & 0.9450 & 0.6091 & 0.6498 & 0.8890 \\
      & Zero-Shot LLM  & 0.1746 & 0.3522 & 0.3026 & 0.1135 & 0.1594 & 0.3905 & 0.3012 & 0.3557 & 0.6764 & 0.2504 & 0.3579 & \second{0.9714} & 0.6687 & 0.7069 & 0.9417 \\
      & Few-Shot LLM   & \best{0.1779} & \second{0.3606} & \second{0.3132} & \second{0.1177} & \second{0.1647} & \second{0.3972} & \second{0.3028} & \second{0.3579} & \second{0.6803} & \second{0.2510} & \second{0.3588} & 0.9707 & \second{0.6746} & \second{0.7109} & 0.9450\\
      & \textbf{Doc2Query++ (Ours)}  & \second{0.1775} & \best{0.3612} & \best{0.3212} & \best{0.1212} & \best{0.1726} & \best{0.4141} & \best{0.3123} & \best{0.3693} & \best{0.6818} & \best{0.2512} & \best{0.3612} & \best{0.9736} & \best{0.6913} & \best{0.7289} & \best{0.9583}\\
    \bottomrule
  \end{tabular}
  }
  \caption{Retrieval performance of different document expansion methods on NFCorpus, SCIDOCS, FiQA-2018, Arguana, and Scifact for both sparse (BM25) and Dual-Index Fusion dense (Contriever) retrieval. Best results are shown in \textbf{bold}, and second-best are \underline{underlined}.}
  \label{tab:main-results}
\end{table*}

\label{dataset-stats}

\textbf{Datasets.} To assess the generalization of DE methods across diverse domains and retrieval tasks, we evaluate on BEIR \cite{thakur2021beir}, a heterogeneous benchmark aggregating 18 public datasets spanning news, science, finance, social media, and more. Unlike prior work focused on in-domain collections like MSMARCO \cite{bajaj2016ms} or TREC-CAR \cite{dietz2017trec}, BEIR emphasizes out-of-domain transfer, making it ideal for testing cross-domain robustness. We select five representative subsets from BEIR to ensure diversity in domains (biomedical, scientific literature, finance, argumentation, and fact-checking), task types (e.g., full-text retrieval, citation prediction, claim verification), and corpus sizes:
\begin{itemize}
    \item NFCorpus~\cite{boteva2016full}: Biomedical retrieval with 323 natural language queries from NutritionFacts.org matched to 3,633 PubMed abstracts.
    \item SCIDOCS~\cite{cohan2020specter}: Citation prediction framed as retrieval; 1,000 paper abstracts as queries over 25,657 scientific articles.
    \item FiQA-2018~\cite{maia201818}: Financial QA with 648 investor questions over 57,638 passages from news and reports.
    \item ArguAna~\cite{wachsmuth2018retrieval}: Counter-argument retrieval; 1,406 debate arguments as queries over 8,674 snippets.
    \item SciFact~\cite{wadden2020fact}: Scientific claim verification with 300 claims over 5,183 abstracts.
\end{itemize}

\noindent \textbf{Metrics.} Following BEIR, we employ nDCG@10 (N@10) and Recall@100 (R@100) to evaluate retrieval effectiveness. N@10 weights early-ranked relevant documents higher, normalized for comparability. R@100 measures the proportion of relevant documents in the top 100 results. We also report MAP to balance precision and recall across the full ranked list.

\noindent \textbf{Retrieval:} For sparse retrieval, we employ BM25~\cite{robertson1994bm25} via PyTerrier~\cite{pyterrier2020ictir} with default parameters ($k_1=0.9$, $b=0.4$). Expanded documents are formed by appending generated queries to the original text before indexing. For dense retrieval, we use Contriever~\cite{izacard2021unsupervised}, an unsupervised dual-encoder pre-trained on large corpora for strong zero-shot performance on BEIR. We encode passages with the \texttt{facebook/contriever-msmarco} checkpoint and index embeddings in FAISS~\cite{douze2024faiss} for inner-product search. To mitigate noise from query concatenation, we apply our Dual-Index Fusion strategy (explained in Section \ref{dual-index-method}) with $\alpha=0.5$, $n_t = 300$ and $n_q=1000$.

\noindent \textbf{Baselines:} We compare Doc2Query$++$ against several document expansion approaches. For fair comparison, all methods generate 30 queries per document. LLM-based methods (including ours) use LLaMA-3.1-8B-Instruct \cite{grattafiori2024llama} with temperature 0.8 to generate 3 queries per prompt until 30 queries are reached.

\begin{itemize}
    \item Doc2Query \cite{nogueira2019document}: popular document expansion technique using seq2seq model trained on MS MARCO to generate synthetic queries and append them to the original document for lexical search. We replicate the setup of Nogueira and Lin \cite{nogueira2019document} to use T5-base model and apply top-$k$ sampling ($k=10$).
    \item Doc2Query$--$ \cite{gospodinov2023doc2query}:  Extends Doc2Query with ELECTRA \cite{clark2020electra}-based filtering for low-relevance queries. We generate an initial pool of 100 via Doc2Query, then retain the top 30\% ($\sim$30) by cross-encoder scores.
    \item Zero-shot LLM: As a necessary baseline—since our work uses an LLM rather than T5-we prompt the LLM with the document to generate queries without examples or filtering.
    \item Few-shot LLM: Inspired by Promptagator \cite{dai2022promptagator}, which demonstrates significant performance gain on 11 BEIR \cite{thakur2021beir} tasks by steering the LLM toward task-specific query distributions, our few-shot baseline adopts the same setup, supplying 6 query-document example pairs in the prompt.
\end{itemize}

\noindent \textbf{Implementation Details:} All experiments ran on three NVIDIA RTX A5000 GPUs. We host LLaMA-3.1-8B-Inst \cite{grattafiori2024llama} locally using vLLM \cite{kwon2023efficient}, which supports all our prompt-based stages: topic representation fine-tuning, keyword selection, and query generation. For topic modeling, we apply BERTopic \cite{grootendorst2022bertopic} with default settings. Keyword extraction employs KeyBERT \cite{grootendorst2020keybert} with defaults, except enabling \texttt{use\_mmr=true} for non‐redundant keyword selection. Both rely on domain‐specific SBERT encoders: BioBERT \cite{deka2022evidence} for NFCorpus, SPECTER \cite{specter2020cohan} for SCIDOCS, all-mpnet-v2 \cite{reimers2021allmpnetbasev2} for Scifact and Arguana, and Fin-MPNET \cite{mukaj_fin_mpnet_base_2024} for FiQA-2018. For few-shot prompting in Doc2Query$++$, we sample 6 held-out examples per dataset, derive topics/keywords via BERTopic/KeyBERT, and manually craft queries to illustrate topic coverage.

\subsection{Main Performance Comparison}
\subsubsection{Effectiveness of Doc2Query$++$}
Table ~\ref{tab:main-results} presents the retrieval performance of various DE approaches across both sparse and dense settings. Doc2Query++ consistently achieves superior retrieval performance, robustly outperforming all baselines across datasets and both sparse and dense retrieval settings. This validates the effectiveness and generalizability of our topic coverage-based Query Generation (QG) strategy.
\subsubsection{Analysis of Doc2Query series}
When comparing Doc2Query and Doc2Query$--$, we find that Doc2Query yields higher nDCG\@10 across all datasets and retrieval types—except for FiQA-2018 under dense retrieval—highlighting that \textbf{Doc2Query$--$'s filtering mechanism is largely in-domain specific and struggles to generalize}. In contrast, Doc2Query retains more diverse and dataset-relevant terms, better supporting domain-sensitive information retrieval.
\subsubsection{Analysis Across Retrieval Paradigms}
Analysis across retrieval paradigms reveals a clear trend: LLM-based approaches generally outperform Doc2Query series in the dense setting, suggesting dense retrievers benefit more from the semantically richer, "human-like" queries produced by LLMs over the lexical keyword focus of Doc2Query series. In the sparse setting, Doc2Query series performance is mixed but often competitive with LLMs, suggesting that their emphasis on keyword extraction remains highly useful in the lexical overlap comparison required by sparse retrieval. Crucially, our LLM-based Doc2Query$++$ maintains its lead in both settings, confirming that its keyword selection and topic coverage process effectively ensures sufficient lexical specificity for strong sparse retrieval without sacrificing semantic richness. Furthermore, Few-shot LLM techniques consistently outperform Zero-shot in dense retrieval. This strongly suggests that dense retrievers benefit from in-context learning \cite{dai2022promptagator}, which better aligns the generated query semantics with the target dataset's characteristics, a benefit that is not that critical in the less-sensitive sparse setting.
\begin{table}[t]
  \centering
  \small
  \begin{tabular}{
      l
      S[table-format=1.3] S[table-format=1.3]
      S[table-format=1.3] S[table-format=1.3]
      S[table-format=1.3] S[table-format=1.3]
  }
    \toprule
    \multirow{2}{*}{Dataset}
      & \multicolumn{2}{c}{F}
      & \multicolumn{2}{c}{F+K}
      & \multicolumn{2}{c}{Doc2Query$++$} \\
    \cmidrule(lr){2-3} \cmidrule(lr){4-5} \cmidrule(lr){6-7}
      & {N@10} & {R@100}
      & {N@10} & {R@100}
      & {N@10} & {R@100} \\
    \midrule
    FiQA-2018
      & \num{0.2789} & \num{0.5869}
      & \underline{0.2854} & \underline{0.5888}
      & \textbf{0.2972} & \textbf{0.6197} \\
    SCIDOCS
      & \num{0.1406} & \num{0.3323}
      & \underline{0.1461} & \underline{0.3664}
      & \textbf{0.1568} & \textbf{0.3749} \\
    NFCorpus
      & \num{0.3371} & \underline{0.2531}
      & \underline{0.3374} & \num{0.2519}
      & \textbf{0.3415} & \textbf{0.2639} \\
    Arguana
      & \num{0.3304} & \num{0.9465}
      & \underline{0.3488} & \underline{0.9622}
      & \textbf{0.3561} & \textbf{0.9629} \\
    Scifact
      & \underline{0.6888} & \num{0.9194}
      & \num{0.6871} & \underline{0.9211}
      & \textbf{0.6945} & \textbf{0.9248} \\
    \bottomrule
  \end{tabular}
  \caption{Impact of Keyword and Topic Guidance on Sparse Retrieval. We report nDCG@10 (N@10) and Recall@100 (R@100) for three incremental configurations: F (Few-shot baseline), F+K (F with Keyword Guidance), and Doc2Query$++$ (F+K with Topic Coverage Guidance). The results confirm the complementary value of both constraints.}
  \label{tab:ablation}
\end{table}

\subsection{Study of Doc2Query$++$}
\subsubsection{Ablation Study}
We perform systematic ablation studies to assess the individual contributions of each core module within our proposed method under sparse retrieval. Specifically, we evaluate the retrieval effectiveness across three incremental configurations: \textbf{(1) Few-shot Prompting (F)}, which establishes a baseline using a basic prompt format devoid of external constraints; \textbf{(2) Few-shot + Keyword Guidance (F+K)}, which enhances the base prompt by incorporating extracted document keywords to lexically anchor the generation; and finally, our full method, \textbf{(3) Doc2Query++}, which extends F+K by integrating Topic Coverage Guidance (derived via BERTopic) to encourage the generation of queries that cover high-level semantic aspects. As shown in Table \ref{tab:ablation}, we observe a consistent and monotonic improvement in retrieval performance from F to F+K to Doc2Query++ across all evaluated datasets. This clear trend supports that both keyword guidance and topic guidance provide complementary and positive contributions to the effectiveness and utility of the generated queries for boosting retrieval performance.

\subsubsection{Effectiveness of Topic Coverage}
To better understand the contribution of topic coverage, we analyze how well the generated queries cover the topical content of their source documents and how this affects retrieval performance. Specifically, we compute the correlation between \textbf{topic recall} and \textbf{retrieval performance gain}, measured as nDCG@10 and Recall@100 improvements over the backbone retriever. We measure topic recall by computing the overlap between a document's gold topics and the topics inferred from its generated queries. Each query is embedded and compared to topic centroids (from BERTopic) using cosine similarity. We hard-assign each query to its nearest centroids, and define recall as:
\begin{equation}
\text{Topic Recall} = \frac{|\text{Assigned Topics} \cap \text{Gold Topics}|}{|\text{Gold Topics}|}
\end{equation}

% \begin{figure}[!htbp]
%     \centering
%     \begin{subfigure}[t]{0.8\linewidth}
%         \centering
%         \includegraphics[width=\linewidth]{figures/topic_recall_impact_SCIDOCS.pdf}
%         \label{fig:topic-recall-scidocs}
%     \end{subfigure}
%     \hfill
%     \begin{subfigure}[t]{0.8\linewidth}
%         \centering
%         \includegraphics[width=\linewidth]{figures/topic_recall_impact_FiQA-2018.pdf}
%         \label{fig:topic-recall-fiqa}
%     \end{subfigure}
%     \caption{Relationship between topic recall and retrieval performance gain across datasets. Higher topic recall consistently leads to greater nDCG@10 improvement over the backbone retrievers (BM25 for sparse, Contriever for dense) on FiQA-2018. On SCIDOCS, the correlation holds in dense retrieval but is weak in sparse retrieval, likely due to the dataset’s reliance on domain-specific terminology and lexical matching.}
%     \label{fig:topic-recall-all}
% \end{figure}

\begin{figure}[!htbp]
    \centering
    \includegraphics[width=\linewidth]{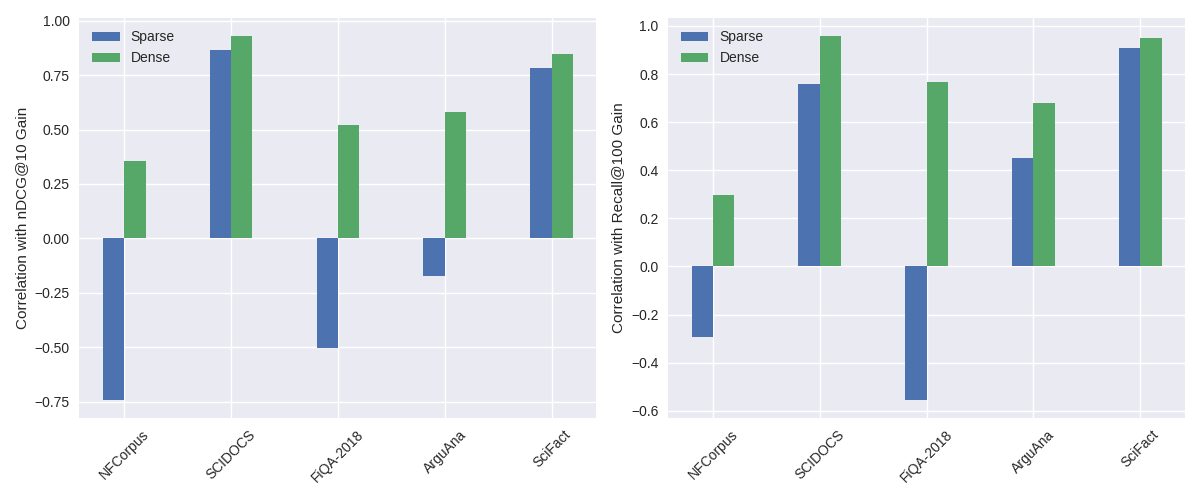}
    \caption{Pearson Correlation between topic recall and Retrieval Gain}
    \label{fig:topic-recall-effectiveness}
\end{figure} 
As shown in Figure ~\ref{fig:topic-recall-effectiveness}, across BEIR subsets, topic recall correlates strongly and positively with performance gains in dense retrieval, as topic coverage enriches embeddings with diverse semantic signals—novel terms from underrepresented facets integrate smoothly via vector averaging in models like Contriever, enhancing latent matches without precision loss. In sparse retrieval, however, correlations weaken or turn negative. We attribute this to BM25's term-frequency weighting amplifying mismatches from coverage-induced novel vocabulary, diluting original document signals in the inverted index and favoring exact overlaps over broad expansions—a shortfall prior methods exacerbate through uncontrolled generation. Nonetheless, topic recall underscores Doc2Query$++$'s value in delivering complete expansions, with our Dual-Index Fusion mitigating sparse noise by isolating query contributions, yielding balanced gains across paradigms.

\subsubsection{Effectiveness of LLM Keyword Selection}
\label{llm-keyword-select-effectiveness}
\begin{table}[t]
  \centering
  \small
  \begin{tabular}{
      l
      S[table-format=1.3] S[table-format=1.3]
      S[table-format=1.3] S[table-format=1.3]
      S[table-format=1.3] S[table-format=1.3]
  }
    \toprule
    \multirow{2}{*}{Dataset}
      & \multicolumn{2}{c}{KeyBERT}
      & \multicolumn{2}{c}{Topic}
      & \multicolumn{2}{c}{LLM} \\
    \cmidrule(lr){2-3} \cmidrule(lr){4-5} \cmidrule(lr){6-7}
      & {N@10} & {R@100}
      & {N@10} & {R@100}
      & {N@10} & {R@100} \\
    \midrule
    FiQA-2018
      & \textbf{0.2663} & \num{0.5623}
      & \num{0.2558} & \underline{0.5690}
      & \underline{0.2619} & \textbf{0.5692} \\
    SCIDOCS
      & \underline{0.1433} & \num{0.3449}
      & \num{0.1323} & \underline{0.3475}
      & \textbf{0.1437} & \textbf{0.3569} \\
    NFCorpus
      & \underline{0.3172} & \num{0.2427}
      & \num{0.3141} & \textbf{0.2485}
      & \textbf{0.3253} & \underline{0.2464} \\
    Arguana
      & \num{0.3462} & \num{0.9450}
      & \underline{0.3550} & \underline{0.9493}
      & \textbf{0.3554} & \textbf{0.9557} \\
    Scifact
      & \underline{0.6791} & \underline{0.9192}
      & \num{0.6667} & \num{0.9107}
      & \textbf{0.6818} & \textbf{0.9223} \\
    \midrule
    Avg.
      & \underline{0.35042} & \num{0.60282}
      & \num{0.34478} & \underline{0.605}
      & \textbf{0.35314} & \textbf{0.61066} \\
    \bottomrule
  \end{tabular}
  \caption{Results on five datasets comparing different Keywords Extraction strategy.}
  \label{tab:kw_comp}
\end{table}

To assess the impact of our hybrid keyword selection strategy, we compare three keyword sources across BEIR subsets: \textbf{KeyBERT} (document-level keyword extraction), \textbf{Topic} (aggregated keywords from BERTopic's c-TF-IDF per topic), and our \textbf{LLM}-based method, which synthesizes both sources for a balanced keyword set. For each, we append the top-10 selected keywords to the original document before indexing, measuring nDCG@10 (precision-oriented) and Recall@100 (coverage-oriented) on sparse retrieval (BM25) performance.

As shown in Table~\ref{tab:kw_comp}, our LLM-based selection yields the highest average scores in both nDCG@10 (0.35314) and Recall@100 (0.61066), outperforming the next best average scores in both precision and recall metrics. These results also clearly validate the underlying motivations of our hybrid strategy by confirming the \textbf{Precision vs. Recall trade-off} in keyword sources: \textbf{KeyBERT} generally outperforms \textbf{Topic} in nDCG@10 (precision), as document-grounded terms reinforce existing signal. Conversely, \textbf{Topic}-based selection outperforms \textbf{KeyBERT} in Recall@100 on most datasets (4 out of 5), demonstrating that topic-level terms introduce necessary diversity and broader vocabulary to boost coverage. Our \textbf{LLM selection successfully synthesizes these complementary strengths}, generating a balanced final keyword set that establishes strong guidance for the subsequent topic-coverage-guided query generation framework.
\begin{table}[t]
  \centering
  \footnotesize
  \setlength{\tabcolsep}{3pt} % reduce horizontal padding
  \begin{tabular}{l cc cc cc cc}
    \toprule
    \multirow{2}{*}{Dataset}
      & \multicolumn{2}{c}{No Expansion}
      & \multicolumn{2}{c}{Append}
      & \multicolumn{2}{c}{Dual-Index}
      & \multicolumn{2}{c}{Gain (\%)} \\
    \cmidrule(lr){2-3} \cmidrule(lr){4-5} \cmidrule(lr){6-7} \cmidrule(lr){8-9}
      & N@10 & R@100 
      & N@10 & R@100 
      & N@10 & R@100 
      & N@10 & R@100 \\
    \midrule
    NFCorpus & 0.331 & \underline{0.295} & \underline{0.332} & 0.294 & \textbf{0.361} & \textbf{0.321} & 8.9 & 8.8 \\
    SCIDOCS & 0.146 & 0.374  & \underline{0.154} & \underline{0.399}  & \textbf{0.173} & \textbf{0.414} & 12.1 & 3.8 \\
    FiQA-2018 & \underline{0.332} & 0.656 & 0.322 & \underline{0.669} & \textbf{0.369} & \textbf{0.682} & 11.1 & 1.9 \\
    Arguana & 0.330 & 0.956 & \underline{0.352} & \underline{0.971} & \textbf{0.361} & \textbf{0.974} & 2.6 &  0.2\\
    Scifact & \underline{0.657} & \underline{0.947} & 0.656 & 0.917 & \textbf{0.729} & \textbf{0.958} & 10.9 &  1.2\\
    \bottomrule
  \end{tabular}
  \caption{Performance comparison between No-Expansion, Append, and Dual-Index strategies across five datasets. Gain is computed as values of \textit{Dual-Index} - \textit{Second Best})/\textit{Second Best} in percentage.}
  \label{tab:comp_dual_append}
\end{table}

\subsubsection{Effectiveness of Dual-Index Fusion}
\label{sec:dual-index-effectiveness}
To assess the effectiveness of our proposed Dual-Index Fusion strategy for dense retrieval, we compare two dense indexing strategies for each document expansion method: the \textbf{Append} strategy, which is identical to sparse retrieval document expansion where generated queries are simply concatenated to the original text before encoding, and our proposed \textbf{Dual-Index Fusion} strategy (Section \ref{dual-index-method}), which encodes the original text and generated queries independently and aggregates their scores via weighted fusion at retrieval time. As shown in Table~\ref{tab:comp_dual_append}, the Dual-Index Fusion approach consistently outperforms the simpler Append strategy across all evaluated QG methods, datasets, and metrics. These consistent improvements align with prior work suggesting that naively appending generated queries often degrades strong dense bi-encoders by integrating excessive semantic noise into the original document embeddings \cite{weller2023generative}. In contrast, Dual-Index Fusion enables more robust retrieval by preserving both the original document and the generated query information independently before aggregating their scores.
\subsubsection{Impact of Fusion Weight on Retrieval Performance}
We conducted an analysis of the fusion weight \(\alpha\) in our Dual-Index Fusion strategy, revealing its critical role in balancing text and query contributions, with detailed results in Figure~\ref{fig:alpha-effectiveness} (Appendix). Precision (nDCG@10) peaks at \(\alpha \approx 0.3\)--\(0.5\) for FiQA and \(\alpha \approx 0.5\)--\(0.6\) for SCIDOCS, declining as query dominance (\(\alpha \to 1.0\)) undermines ranking accuracy. Recall rises with \(\alpha\), enhancing coverage, but all methods drop at \(\alpha = 1.0\), highlighting text embeddings' essential semantic depth. This positions \(\alpha\) as a precision--recall tuner, with optimal values (\(\simeq 0.3\)--\(0.6\)) maximizing ranking quality and higher settings boosting recall, suggesting potential for dynamic optimization. Notably, Doc2Query and Doc2Query-- show recall declines for \(\alpha > 0\), underperforming text-only baselines due to their expansions' limited semantic diversity.

\subsubsection{Effect of Query Quantity on Retrieval Performance}
\begin{figure}[!htbp]
    \centering
    \includegraphics[width=0.9\linewidth]{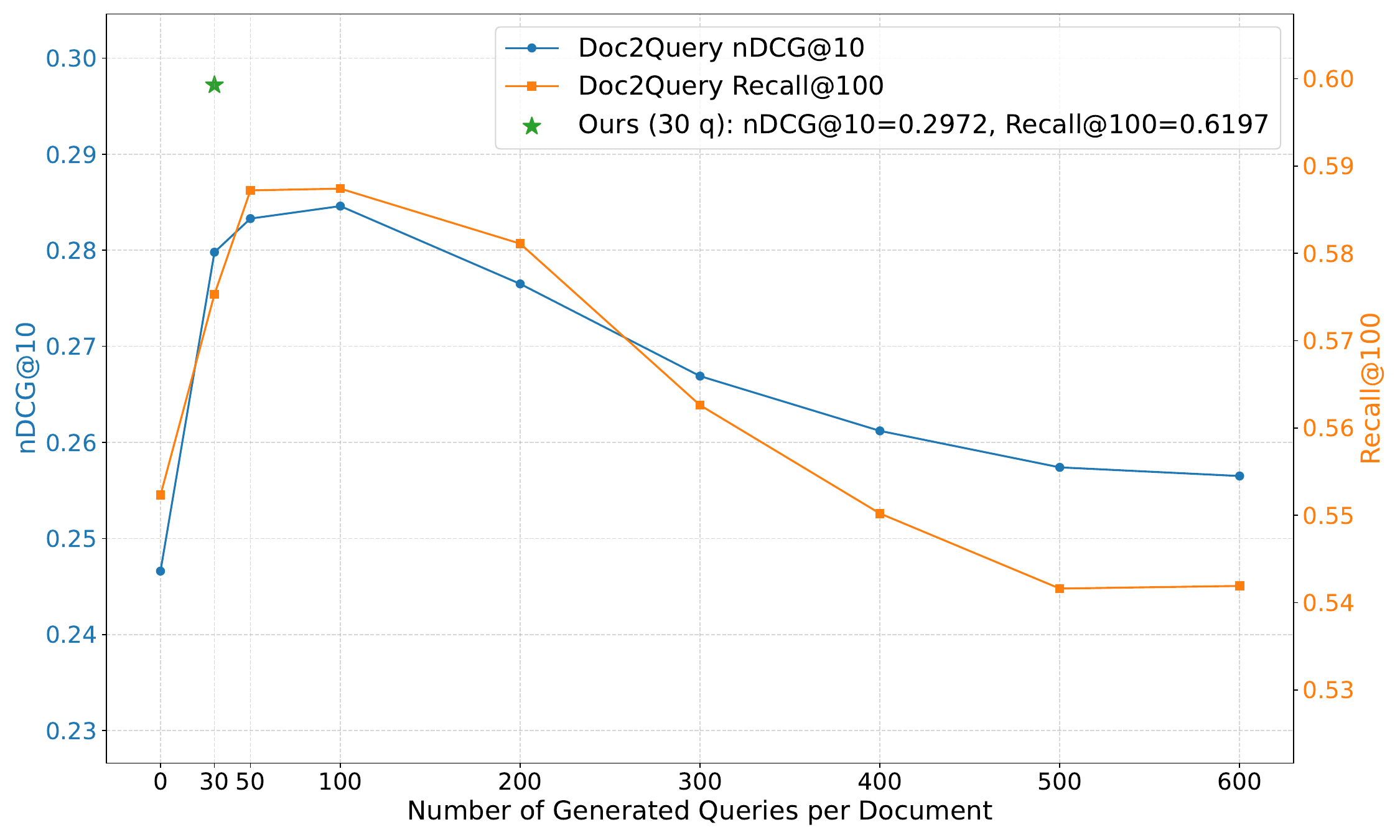}
    \caption{Retrieval performance (nDCG@10 and Recall@100) on FiQA-2018 with varying numbers of generated queries using Doc2Query. Performance peaks around 100 queries and degrades afterward, while our coverage-aware method (30 queries) outperforms even the best Doc2Query setting.}
    \label{fig:query-scaling}
\end{figure}

A key research question we sought to answer is: Why do we need to consider topic coverage when generating queries? Wouldn't generating more queries eventually cover all aspects of a document? To investigate this, we designed an experiment using the FiQA-2018 dataset under sparse retrieval, scaling the number of Doc2Query-generated queries per document from 0 up to 600 (e.g., 30, 50, 100, 200, etc.). As shown in Figure~\ref{fig:query-scaling}, both nDCG@10 and Recall@100 increase as more queries are added, peaking at around 100 queries. However, beyond that point, performance drops significantly, revealing a saturation point where excessive queries introduce redundancy or noise rather than improving coverage. In contrast, Doc2Query$++$ proves that effectiveness is a function of topical breadth, not just query quantity. Our method uses only 30 queries per document, yet outperforms Doc2Query at its optimal point (nDCG@10: 0.2972 vs. 0.2902; Recall@100: 0.6197 vs. 0.5981). Naive scaling cannot substitute for structured, coverage-aware query generation.

\section{Conclusion}
This work revisits document expansion in the era of LLM, arguing that \textit{coverage, not quantity}, is the missing dimension in neural expansion. Through Doc2Query$++$, we demonstrate that structuring query generation around a document’s latent topics fundamentally changes how expansions contribute to retrieval — turning a previously ad-hoc process into one that is interpretable, controllable, and generalizable across domains. Our findings yield several key takeaways for the IR community: (1) Coverage-guided expansion consistently outperforms uncontrolled generation, confirming that diversity and completeness drive retrieval gains; (2) The proposed Dual-Index Fusion reframes expansion for dense models, showing that decoupling text and generated queries converts what was once harmful noise into a complementary retrieval signal; and (3) when guided by topic and keyword, LLMs can act as structured knowledge expanders, suggesting a path toward controllable and domain-agnostic IR augmentation. Beyond performance, Doc2Query$++$ highlights an important conceptual shift: document expansion need not rely solely on data-driven fine-tuning but can instead be architecturally constrained to ensure semantic completeness. This principle invites future research on (1) adaptive topic control mechanisms that dynamically adjust expansion according to document or query characteristics and (2) retrieval-informed feedback loops that iteratively refine topic coverage based on observed retrieval gaps. Also, extending these ideas to multilingual and cross-modal settings further offers a compelling test of Doc2Query++’s generality, pushing toward a generation framework that is not only effective but also adaptive and interpretable across retrieval paradigms.
\bibliographystyle{ACM-Reference-Format}
\bibliography{references}

\appendix
% \section{Fusion Weight Effectiveness}
\begin{figure}[h]
  \centering
  \begin{subfigure}[h]{\linewidth}
    \centering
    \includegraphics[width=0.9\linewidth]{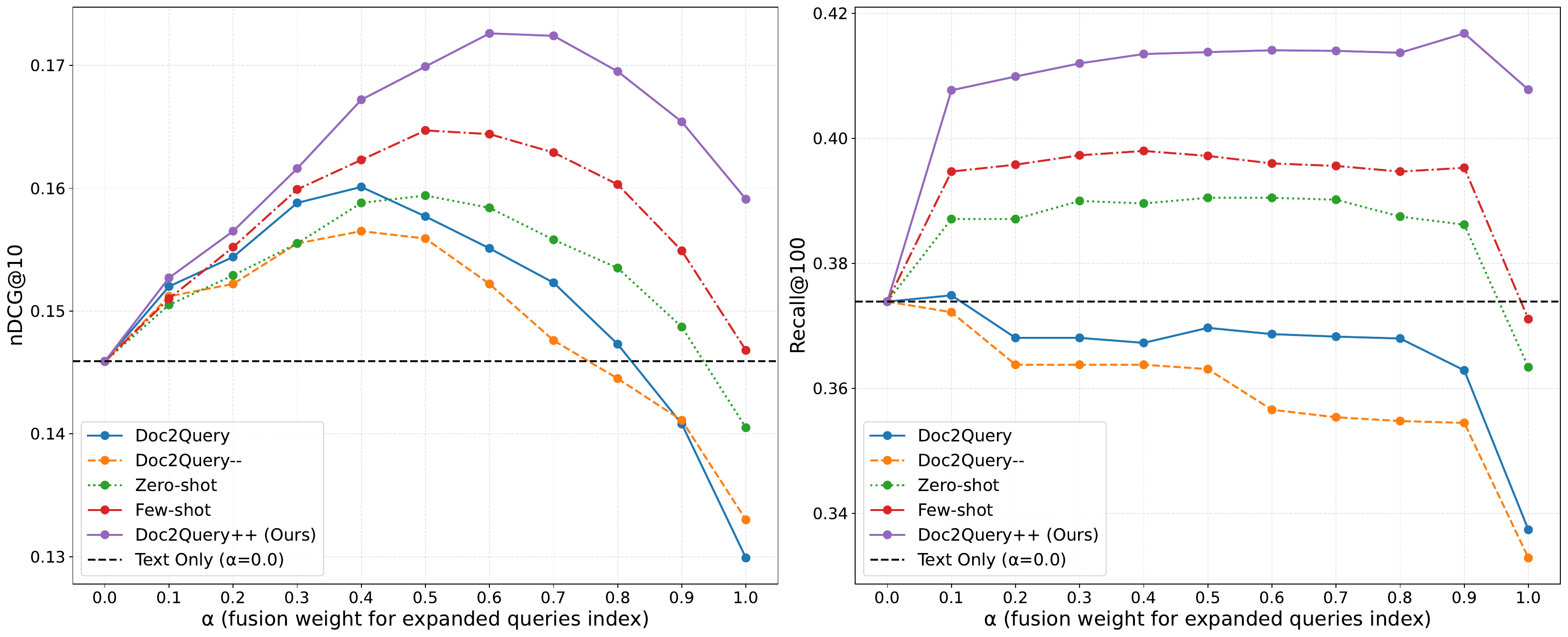}
    \caption{SCIDOCS}
    \label{fig:scidocs-metric-alpha}
  \end{subfigure}
  \vskip 0.5em
  \begin{subfigure}[h]{\linewidth}
    \centering
    \includegraphics[width=0.9\linewidth]{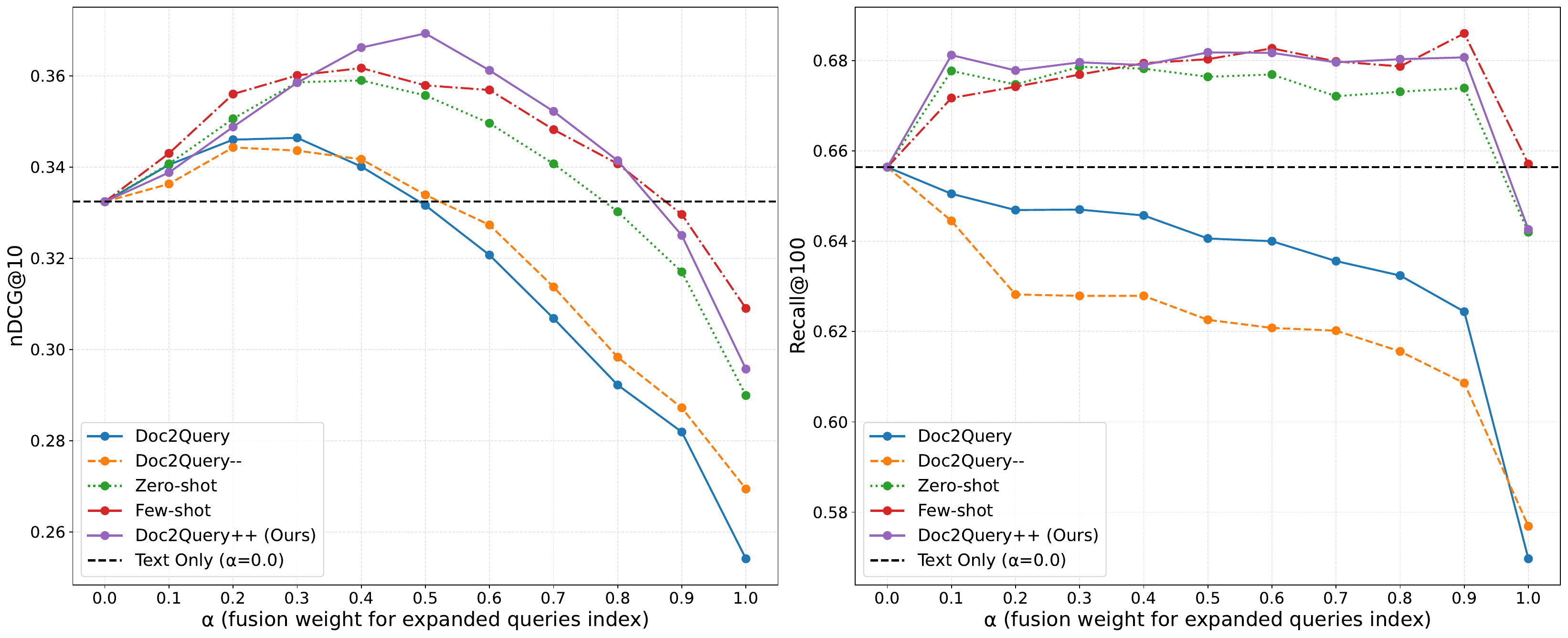}
    \caption{FiQA-2018}
    \label{fig:fiqa-metric-alpha}
  \end{subfigure}
  \caption{Impact of fusion weight $\alpha$ on retrieval performance across different document expansion methods.}
  \label{fig:alpha-effectiveness}
\end{figure}

% \section{Statistics of Captured Topics}
% \label{apdix:topic_stats}
% In Table~\ref{tab:topic_stats}, we report the total topic number captured by BERTTopic and the average topic number of each document across five datasets.
% \input{tables/topic_stats}

\section{Full Prompts for Query Generation}
Here, we provide all the prompts used in Doc2Query++, including prompts for topic name refinement (described in Section~\ref{sec:topic-refine-llm}), Keywords Selection (describe in Section~\ref{sec:keywords_extract_select}), and Topic-Coverage Multi-Query Generation (described in Section~\ref{method:qg-stage}).\footnote{Furthermore, we provide the fewshot examples of all the datasets used in topic refinement and topic-coverage multi-query generation in the anonymous github. \url{https://anonymous.4open.science/r/doc2queryPlusPlus-41BB/}}
\subsection{Prompt for Topic Name Refinement}
\label{topic-name-refinement-prompt}
\begin{tcolorbox}[
  breakable,
  enhanced,
  colback=white,
  colframe=blue!50!black,
  title=Few-shot Template for Natural Language Topic Name Generation,
  fontupper=\footnotesize,
  coltitle=black,
  colbacktitle=blue!10,
]

You will extract a short topic label from given documents and keywords.
Here are four examples of topics you created before:

\vspace{0.5em}
\textbf{Example 1}

\textbf{Sample texts from this topic:}
\begin{itemize}[leftmargin=*]
  \item But if you believe the price will go down, the only way to buy low and sell high is to sell first and buy later.
  \item With your comment, you have stated that your scenario is that you believe that the stock will go up long term, but you also believe that the stock is at a short-term peak and will drop in the near future.
  \item You believe that the stock is a long-term buy, but for some reason you are guessing that the stock will drop in the short-term.
\end{itemize}

\textbf{Keywords:} stock, the stock price, stock price, stock is, the price, buy, sell, value, share

\textbf{Topic:} Short-Term Stock Trading

\vspace{0.5em}
\textbf{Example 2}\\
...\\
\textbf{Example 3}\\
...\\
\textbf{Example 4}\\
...\\

\vspace{1em}
\textbf{Your Task}

\textbf{Sample texts from this topic:}
\texttt{[DOCUMENTS]}

\textbf{Keywords:}
\texttt{[KEYWORDS]}
\\
\textbf{**Crucial Output Instruction:**\\
You MUST generate a single line as your response.\\
This line MUST start EXACTLY with `topic: ` (including the space after the colon).\\
Following `topic: `, provide ONLY the concise topic label.\\
Do NOT add any other text, explanations, numbering, markdown, or any content before or after this single line.}

\textbf{Topic:}
\end{tcolorbox}
\subsection{Prompt for Keywords Selection}
\label{keyword-selection-prompt}
\begin{tcolorbox}[
  breakable,
  enhanced,
  colback=white,
  colframe=blue!50!black,
  title=Prompt Template for Keyword Selection,
  fontupper=\footnotesize,
  coltitle=black,
  colbacktitle=blue!10,
]

You will receive a document along with a set of candidate keywords. Your task is to select the keywords that best align with the core theme of the document. Exclude keywords that are too broad or less relevant. You may list up to [final keyword num] keywords, using only the keywords in the candidate keyword set:

\textbf{Document:}
[DOCUMENT]

\textbf{Candidate keyword set:}
[Topic-Level Keywords + Document-Level Keywords]

\textbf{Final Keywords:}
\end{tcolorbox}
\subsection{Prompt for Topic-Coverage Multi-Query Generation}
\label{few-shot-prompt-qg}
\begin{tcolorbox}[
  breakable,
  enhanced,
  colback=white,
  colframe=blue!50!black,
  title=Few-shot Template for Multi-Query Generation from Document and Keywords,
  fontupper=\footnotesize,
  coltitle=black,
  colbacktitle=blue!10,
]

You are an expert assistant in crafting search queries for a given passage that cover specified topics and make use of given keywords. The following are some examples:

\vspace{0.5em}
\textbf{Example 1}

\textbf{Article:}
You have a good thing going. One of the luxuries of being invested in an index fund for the long term is that you don't have to sweat the inevitable short term dips in the market. Instead, look at the opportunity that presents itself on market dips: now your monthly investment is getting in at a lower price. Buy low, sell high. Don't lose money. These are common mantras for long term investment mentality. 5-8 years is plenty of time -- I'd call it medium-term. As you get closer to your goals (2-3 years out) you should start slowly moving money out of your index fund and start dollar cost averaging out into cash or short-term bonds (but that's another question). Keep putting money in, wait, and sell high. If it's not high, wait another year or two to buy the house. A lot of people do the opposite for their entire lives: buying high, panic selling on the dips, then buying again when it goes up. That's bad! I recommend a search on dollar cost averaging, which is exactly what you are doing right now with your monthly investments.

\textbf{Topics:}
Index-fund investing, Dollar-cost averaging, Investment horizon planning

\textbf{Keywords:} index fund, long-term investment, market dips, "buy low, sell high", "don't lose money", medium-term horizon (5–8 years), cash/short-term bonds, panic selling, monthly contributions, dollar cost averaging

\textbf{Generated Queries:}
\begin{itemize}[leftmargin=*]
  \item How does dollar cost averaging help with index fund investing during market dips?
  \item When should I transition from index funds to cash or short-term bonds for medium-term goals?
  \item What are the benefits of long-term investment strategy versus panic selling during market downturns?
\end{itemize}

\vspace{0.5em}
\textbf{Example 2}\\
...\\
\textbf{Example 3}\\
...\\
\textbf{Example 4}\\
...\\
\textbf{Example 5}\\
...\\
\textbf{Example 6}\\
...\\

\vspace{1em}
\textbf{Your Task:}

Now generate 3 relevant queries for this passage that collectively cover specified topics by using given keywords:

\textbf{Passage:} \texttt{<document $d$>}

\textbf{Topics:} \texttt{<topics $T_d$>}

\textbf{Keywords:} \texttt{<keywords $K_d$>}

\textbf{Queries:}
\end{tcolorbox}

\end{document}